\documentclass[prl,twocolumn,showpacs,superscriptaddress]{revtex4}

\usepackage{amsmath,amssymb}
\usepackage{verbatim}
\usepackage{graphicx}
\usepackage[
      colorlinks=true,
      linkcolor=blue,
      urlcolor=magenta,
      filecolor=green,
      citecolor=red,
      pdfstartview=FitV,
      pdftitle={},
        pdfauthor={Hamid Afshar, Stephane Detournay, Daniel Grumiller, Wout Merbis, Alfredo Perez, David Tempo and Ricardo Troncoso},
        pdfsubject={},
        pdfkeywords={},
        pdfpagemode=None,
        bookmarksopen=true
      ]{hyperref}
\usepackage{color}

\DeclareFontFamily{OT1}{rsfs}{}
\DeclareFontShape{OT1}{rsfs}{m}{n}{ <-7> rsfs5 <7-10> rsfs7 <10->rsfs10}{} 
\DeclareMathAlphabet{\mycal}{OT1}{rsfs}{m}{n}

\newcommand{\be}[1]{ \begin{equation}\label{#1} }
\newcommand{\ee}{\end{equation}}
\newcommand{\bea}[1]{\begin{eqnarray}\label{#1} }
\newcommand{\eea}{\end{eqnarray}}

\newcommand{\eq}[2]{\begin{equation} #1 \label{#2} \end{equation}}

\DeclareMathOperator{\extdm}{d}

\newcommand{\extd}{\extdm \!}

\newcommand{\beq}{\begin{equation}}
\newcommand{\eeq}{\end{equation}}
\newcommand{\bi}{\begin{itemize}}
\newcommand{\ei}{\end{itemize}}
\newcommand{\bt}{\begin{tabular}}
\newcommand{\et}{\end{tabular}}
\newcommand{\bc}{\begin{center}}
\newcommand{\ec}{\end{center}}

\def\one{{\hbox{ 1\kern-.8mm l}}}
\newcommand{\Dslash}{\not{\hbox{\kern-4pt $D$}}}
\newcommand{\pdslash}{\not{\hbox{\kern-2pt $\partial$}}}

\newcommand{\ba}{\begin{array}}
\newcommand{\ea}{\end{array}}

\def\bbox{{\,\lower0.9pt\vbox{\hrule \hbox{\vrule height 0.2 cm
\hskip 0.2 cm \vrule height 0.2 cm}\hrule}\,}}
\newcommand{\dsl}{\pa \kern-0.5em /}

\newcommand{\vp}{\varphi}

\begin{document}

\newcommand{\mytitle}{Soft Heisenberg hair on black holes in three dimensions}

\title{\mytitle}

\author{Hamid Afshar}
\email{afshar@ipm.ir}
\affiliation{School of Physics, Institute for Research in Fundamental Sciences (IPM),  P.O.Box 19395-5531, Tehran, Iran}

\author{Stephane Detournay}
\email{sdetourn@ulb.ac.be}
\affiliation{Physique Th\'eorique et Math\'ematique, Universit\'e Libre de Bruxelles and 
International Solvay Institutes, Campus Plaine C.P. 231, B-1050 Bruxelles, Belgium}

\author{Daniel Grumiller}
\email{grumil@hep.itp.tuwien.ac.at}
\affiliation{Institute for Theoretical Physics, TU Wien, Wiedner Hauptstrasse 8--10/136, A-1040 Vienna, Austria}
\affiliation{Centro de Estudios Cient\'i­ficos (CECs), Av. Arturo Prat 514, Valdivia, Chile}

\author{Wout Merbis}
\email{merbis@hep.itp.tuwien.ac.at}
\affiliation{Institute for Theoretical Physics, TU Wien, Wiedner Hauptstrasse 8--10/136, A-1040 Vienna, Austria}

\author{Alfredo Perez}
\email{aperez@cecs.cl}
\affiliation{Centro de Estudios Cient\'i­ficos (CECs), Av. Arturo Prat 514, Valdivia, Chile}

\author{David Tempo}
\email{tempo@cecs.cl}
\affiliation{Centro de Estudios Cient\'i­ficos (CECs), Av. Arturo Prat 514, Valdivia, Chile}

\author{Ricardo Troncoso}
\email{troncoso@cecs.cl}
\affiliation{Centro de Estudios Cient\'i­ficos (CECs), Av. Arturo Prat 514, Valdivia, Chile}

\date{\today}

\preprint{TUW--16--06}

\begin{abstract} 
Three-dimensional Einstein gravity with negative cosmological constant admits stationary black holes that are not necessarily spherically symmetric. 
We propose boundary conditions for the near horizon region of these black holes that lead to a surprisingly simple near horizon symmetry algebra consisting of two affine $\hat{u}(1)$ current algebras. The symmetry algebra is essentially equivalent to the Heisenberg algebra. The associated charges give a specific example of ``soft hair'' on the horizon, as defined by Hawking, Perry and Strominger. We show that soft hair does not contribute to the Bekenstein--Hawking entropy of Ba\~nados--Teitelboim--Zanelli black holes and ``black flower'' generalizations. 
From the near horizon perspective the conformal generators at asymptotic infinity appear as composite operators, which we interpret in the spirit of black hole complementarity. 
Another remarkable feature of our boundary conditions is that they are singled out by requiring that the whole spectrum is compatible with regularity at the horizon, regardless the value of the global charges like mass or angular momentum.  Finally, we address black hole microstates and generalizations to cosmological horizons.
\end{abstract}


\maketitle


\section{Introduction}

The Bekenstein--Hawking (BH) entropy formula for
horizons of area $A$ ($G_{N}$ is Newton's constant)
\begin{equation}
S_{\textrm{\tiny BH}}=\frac{A}{4G_N} 
\label{eq:r1}
\end{equation}
has been a source of inspiration for approaches to quantum gravity
and led to derivations of the entropy \eqref{eq:r1} from a microstate
counting \cite{Carlip:1994gy,Strominger:1996sh,Strominger:1997eq,Carlip:1998wz,Guica:2008mu}.
Many of these approaches exploit either the simplicity of (near-)extremal
black holes, or the power of conformal symmetries, or both. In generic
situations, however, the horizon is non-extremal, not always due to
a black hole, and the near horizon symmetries are not necessarily conformal.

In particular, in flat space both the asymptotic symmetries \cite{Bondi:1962,Sachs:1962,Ashtekar:1996cd,Barnich:2006av}
and [at least in three dimensions (3d)] the near horizon symmetries
\cite{Donnay:2015abr,Afshar:2015wjm} are related with the Bondi--van~der~Burg--Metzner--Sachs
(BMS) algebra \cite{Bondi:1962,Sachs:1962}. The importance of near-horizon
BMS symmetries as a means to understand black holes was recently highlighted
in \cite{Hawking:2016msc}. For related works see \cite{Blau:2015nee,Penna:2015gza,Hooft:2016itl,Bianchi:2016tju,Averin:2016ybl,Compere:2016jwb,Kehagias:2016zry,Compere:2016hzt}.

In this article we explore the spacetime geometry around non-extremal horizons, which is universally approximated by the product of two-dimensional Rindler space \cite{Rindler:1966zz} with a compact Euclidean manifold. For simplicity we shall work in 3d. In a co-rotating frame, the near horizon metric in (ingoing) Eddington--Finkelstein coordinates is given by
\begin{equation}
\extd s^2 = -2ar\,\extd v^2+2\extd v\extd r+\gamma^2\,\extd\varphi^2+\cdots\label{eq:r2}
\end{equation}
where the constant $a$ is the Rindler acceleration. 
The vanishing of the radial coordinate, $r=0$, corresponds to the location of
the horizon, $v$ is the advanced time and we assume periodicity of
the angular coordinate $\varphi\sim\varphi+2\pi$ so that the horizon
is compact and has a total area given by $A=\oint\extd\varphi\,\gamma$.
With no loss of generality we assume $a$ and $\gamma$ to be positive,
and the ellipsis refers to higher order terms in the radial coordinate
$r$ or to rotation terms (we shall be more explicit below).

One of our main goals is to explore the near horizon behaviour of the gravitational field. We shall propose a new set
of boundary conditions consistent with \eqref{eq:r2}, which leads to a very simple near horizon symmetry algebra, the Heisenberg algebra. The associated charges provide a particular manifestation of ``soft hair'' in the sense of \cite{Hawking:2016msc}. We show that the BH entropy is solely determined by the zero-mode charges and does not receive a contribution from the soft hair. 
We then establish how the near horizon symmetries are linked to the ones at infinity \cite{Brown:1986nw,Barnich:2006av}, and interpret our results in the spirit of black hole complementarity \cite{'tHooft:1984re,Susskind:1993if,Stephens:1994an}. We conclude with a discussion of black hole microstates and generalizations to cosmological horizons.




While some of the technical tools available to us are specific to 3d, we believe that the general lessons drawn from our derivations are dimension-independent and thus shed new light on near horizon symmetries, soft hair, microstate counting and black hole complementarity.


\section{Soft hairy black holes}

The behaviour of the gravitational field in 3d general relativity with negative cosmological constant
$\Lambda=-\ell^{-2}$ around a non-extremal horizon can be described by a near horizon metric in ingoing
Eddington--Finkelstein coordinates [$\ell\rho=r$ and $f:=1+\rho/(2a\ell)$]
\begin{multline}
\extd s^{2} =-2a\ell\rho f\extd v^2+2\ell\extd v\extd\rho-2\omega a^{-1}\,\extd\varphi\extd\rho \\
+4\omega\rho f\extd v\extd\varphi+\big[\gamma^2+\tfrac{2\rho}{a\ell} f (\gamma^2-\omega^2)\big]\extd\varphi^2\label{eq:r5}
\end{multline}
where $\omega$ and $\gamma$ are arbitrary functions of $\varphi$. The metric deviates to leading order from \eqref{eq:r2} in the $g_{\rho\varphi}$ component, but this can always be gauged away. It turns out, however, to be convenient keeping it as it is.

The line element (\ref{eq:r5}) is an exact solution of Einstein's equations in 3d, since it has constant curvature. The geometry possesses an event horizon located at $\rho=0$. Since it is not spherically symmetric, the solution generically describes a ``black flower'' \cite{Barnich:2015dvt}.
In the case of constant $\omega$ and $\gamma$ the solution \eqref{eq:r5} reduces
to the Ba\~nados--Teitelboim--Zanelli (BTZ) black hole \cite{Banados:1992wn,Banados:1992gq}. The
metric (\ref{eq:r5}) does not obey Brown-Henneaux boundary
conditions \cite{Brown:1986nw}, which motivates us to propose
boundary conditions that accommodate these solutions. 
This task becomes remarkably simple in the Chern--Simons formulation.


\section{Einstein Gravity as a Chern--Simons theory}

While the metric
formulation is closer to our physical and geometric intuition, the
reformulation of 3d Einstein gravity as Chern--Simons theory is more
powerful at a technical level, which is why we shall use it. The bulk
action reads \cite{Achucarro:1987vz,Witten:1988hc}
\begin{equation}
I_{\text{CS}}=\frac{k}{4\pi}\int\text{\,}\langle{\cal A}\wedge\extd{\cal A}+\tfrac{2}{3}\text{\,}{\cal A}\wedge{\cal A}\wedge{\cal A}\rangle\text{\,}.\label{eq:r3}
\end{equation}
The coupling constant is given by $k=\ell/(4G_{N})$ and the connection
${\cal A}$ decomposes into two $sl(2,\mathbb{R})$ connections $A^{\pm}$
with generators $[L_{n},L_{m}]=(n-m)L_{n+m}$~($n,m=0,\pm1$) such
that the bilinear form $\langle,\rangle$ is essentially the standard
one for each $sl(2,\mathbb{R})$, $\langle L_{1},\, L_{-1}\rangle=-1$,
$\langle L_{\pm1},\, L_{0}\rangle=0$, $\langle L_{0},\, L_{0}\rangle=\tfrac{1}{2}$,
with additional minus signs for $A^{-}$. 

The metric is determined from the connections $A^{\pm}$ as
\begin{equation}
g_{\mu\nu}=\tfrac{\ell^{2}}{2}\left\langle \left(A_{\mu}^{+}-A_{\mu}^{-}\right)\left(A_{\nu}^{+}-A_{\nu}^{-}\right)\right\rangle \,.\label{eq:r8}
\end{equation}
Before we start we list the length dimensions we are using. The quantities
$v,\gamma,\omega,\ell,G_{N}$ have length dimension one, $\rho,\vp,k,A^{\pm},L_{n}$
are dimensionless and Rindler acceleration $a$ has length dimension
minus one.


\section{New boundary conditions}

Based on the near horizon behaviour
of the metric, one is naturally led to propose a new set of boundary
conditions, which in terms of the gauge fields reads
\begin{equation}
A^{\pm}=b_{\pm}^{-1}\big(\extd+\frak{a}^{\pm}\big)b_{\pm}\label{eq:r7}
\end{equation}
where $b_{\pm}=\exp{(\pm\tfrac{1}{\ell\zeta^{\pm}}\, L_{1})} \cdot \exp{(\pm\tfrac{\rho}{2}\, L_{-1})}$.
The auxiliary connection is given by
\begin{equation}
\frak{a}^{\pm}=L_{0}\left(\pm{\cal J}^{\pm}\text{\,}\extd\varphi+\zeta^{\pm}\text{\,}\extd v\right)\label{eq:r21}
\end{equation}
with $\ell{\cal J}^{\pm}:=\gamma\pm\omega$. The state-dependent functions ${\cal J}^{\pm}$ and the (arbitrary but fixed) chemical potentials $\zeta^{\pm}$ (see e.g.~\cite{Compere:2013nba,Henneaux:2013dra}) depend on $\vp$ and $v$ in general. 

The field equations ${\cal F}={\extd}{\cal A}+{\cal A}{\wedge}{\cal A}=0$ hold exactly and yield $\partial_{v}{\cal {J}}^{\pm}=\pm\zeta^{\pm\prime}$, where prime denotes differentiation with respect to $\vp$.

For simplicity we assume constant chemical potentials. Then the dynamical fields
${\cal {J}}^{\pm}$ become independent of the advanced time $v$,
and in the particular case of $\zeta^{\pm}=-a$, from (\ref{eq:r8})
one recovers the spacetime metric (\ref{eq:r5}). 


\section{Canonical generators}

Our next step is to determine the canonical generators $Q[\epsilon^{+},\epsilon^{-}]={\cal Q}^{+}[\epsilon^{+}]-{\cal Q}^{-}[\epsilon^{-}]$
for arbitrary transformations $\epsilon^{\pm}=\epsilon_{i}^{\pm}L_{i}$
that preserve the boundary conditions \eqref{eq:r7}, \eqref{eq:r21}. In the Regge--Teitelboim approach
\cite{Regge:1974zd,Banados:1994tn} their variation is 
\begin{equation}
\delta{\cal Q}^{\pm}[\epsilon^{\pm}]=\mp\frac{k}{4\pi}\oint\extd\varphi\text{\,}\eta^{\pm}\delta{\cal J}^{\pm} \label{eq:rc3}
\end{equation}
with $\eta^{\pm}=\epsilon_{0}^{\pm}$. The most general transformations
$\delta_{\epsilon^{\pm}}\frak{a}^{\pm}=\extd\epsilon^{\pm}+[\frak{a}^{\pm},\,\epsilon^{\pm}]={\cal O}(\delta\frak{a}^{\pm})$
that preserve the boundary conditions \eqref{eq:r21} imply $\delta{\cal {J}}^{\pm}=\pm\eta^{\pm\prime}$,
with $\partial_{v}\eta^{\pm}=0$. The additional components $\epsilon_{\pm1}^{\pm}$ generate trivial
gauge transformations, since they neither appear in the transformation
laws of the dynamical fields nor in the variation of the global charges
\cite{Benguria:1976in}.

The global charges are obtained from functionally integrating \eqref{eq:rc3} and turn out to be finite, integrable and conserved in (advanced) time,
\begin{equation}
{\cal Q}^{\pm}[\eta^{\pm}]=\mp\frac{k}{4\pi}\oint\extd\varphi\text{\,}\eta{}^{\pm}(\varphi)\text{\,}{\cal J}^{\pm}(\varphi)\,.\label{eq:Qmn}
\end{equation}
We highlight that the surface integrals in \eqref{eq:Qmn}
do not depend on the radial coordinate $\rho$, which implies that
the boundary analysis actually holds for any fixed value $\rho=\rho_{0}$,
regardless of whether $\rho_{0}$ is close to the horizon or infinity. As explained
below, this is the key in order to establish the relationship between
near horizon symmetries with the ones at infinity.


\section{Near horizon symmetry algebra}

The algebra of the global charges
captures all boundary condition preserving transformations modulo
trivial gauge transformations. It is determined by the relation $\delta_{\eta_{2}}Q[\eta_{1}]=\{Q[\eta_{1}],\, Q[\eta_{2}]\}$,
where $\{,\}$ denotes Dirac brackets. Expanding in Fourier modes,
$J_{n}^{\pm}=\frac{k}{4\pi}\oint\extd\varphi\, e^{in\varphi}{\cal {J}}^{\pm}\left(\varphi\right)$,
leads to a remarkably simple symmetry algebra
\begin{equation}
\left[J_{n}^{\pm},\, J_{m}^{\pm}\right]=\pm\tfrac{1}{2}kn\delta_{n+m,\text{\,}0}\text{\quad\quad}\left[J_{n}^{+},\, J_{m}^{-}\right]=0\label{eq:NHSA}
\end{equation}
where we made the usual replacement of Dirac brackets by commutators,
$i\{,\}\to[,]$. The algebra \eqref{eq:NHSA} consists of two $\hat{u}(1)$
current algebras with levels $\pm k/2$.

Changing the basis according to $P_{0}=J_{0}^{+}+J_{0}^{-}$,
$P_{n}=\tfrac{i}{kn}\,(J_{-n}^{+}+J_{-n}^{-})$ if $n\neq0$, $X_{n}=J_{n}^{+}-J_{n}^{-}$,
it becomes apparent that the algebra \eqref{eq:NHSA} is equivalent
to the canonical commutation relations for Casimir--Darboux coordinates
(we set $\hbar=1$) 
\begin{align}
\left[X_{n},\, X_{m}\right] & =\left[P_{n},\, P_{m}\right]=\left[X_{0},\, P_{n}\right]=\left[P_{0},\, X_{n}\right]=0 \label{eq:heisenberg1}\\
\left[X_{n},\, P_{m}\right] & =i\delta_{n,m}\quad\textrm{if}\; n\neq0\label{eq:heisenberg}
\end{align}
where $X_{0}$ and $P_{0}$ are the two Casimirs and all other $X_{n},\, P_{n}$
form canonical pairs. Eq.~\eqref{eq:heisenberg} is the Heisenberg
algebra. Thus, we have obtained a surprisingly simple kinematical
Hilbert space. 

The near horizon symmetry algebra \eqref{eq:NHSA} {[}or
equivalently \eqref{eq:heisenberg1}, \eqref{eq:heisenberg}{]} is
a key result of our work.


\section{Soft hair}\label{Sect:softhair}

The dynamics is governed by the Hamiltonian,
whose corresponding surface integral is defined by $H:=Q[\epsilon^{\pm}|_{\partial_{v}}]$,
with $\epsilon^{\pm}|_{\partial_{v}}=\frak{a}_{v}^{\pm}=L_{0}\zeta^{\pm}$ \cite{Witten:1988hc}.
For the particular choice $\zeta^{\pm}=-a$ 
the Hamiltonian is given by
$H=aP_{0}$, 
which commutes with all canonical coordinates $X_{n},\, P_{n}$, so
that we have trivial dynamics.

We consider now all vacuum descendants $|\psi(q)\rangle$ (labelled
by a set $q$ of arbitrary non-negative integer quantum numbers $N^{\pm}$,
$n_{i}^{\pm}$ and $m_{i}^{\pm}$)
\begin{equation}
|\psi(q)\rangle=N(q)\prod\nolimits_{i=1}^{N^{+}}(J_{-n_{i}^{+}}^{+})^{m_{i}^{+}}\prod\nolimits_{i=1}^{N^{-}}(J_{-n_{i}^{-}}^{-})^{m_{i}^{-}}|0\rangle \label{eq:rc40}
\end{equation}
with a normalization constant $N(q)$ such that $\langle\psi(q)|\psi(q)\rangle=1$.
Since $H$ commutes with all generators $J_{n}^{\pm}$ we obtain the
energy of the vacuum, $H|0\rangle=E_{\textrm{\tiny{vac}}}|0\rangle$,
for all descendants $|\psi(q)\rangle$. 
\begin{equation}
E_{\psi}=\langle\psi(q)|H|\psi(q)\rangle=E_{\textrm{\tiny{vac}}}\langle\psi(q)|\psi(q)\rangle=E_{\textrm{\tiny{vac}}}\,.\label{eq:rc39}
\end{equation}

This implies that all descendants of the vacuum have the same energy
as the vacuum, i.e., they are ``soft hair'' in the precise sense
of being zero-energy excitations \cite{Hawking:2016msc}. In the derivation
above we can replace the vacuum state $|0\rangle$ by any other state
with the same conclusions.


\section{Soft hairy black hole entropy}

Choosing constant chemical potentials $\zeta^{\pm}$, the general solution of the field equations with our boundary conditions \eqref{eq:r7},~\eqref{eq:r21} describes a stationary non spherically symmetric black hole that carries all of the possible left and right $\hat{u}(1)$ charges. The simplest case is the spherically symmetric one that corresponds to the BTZ black hole, which only carries zero-mode charges, $J_{0}^{\pm}=\tfrac{1}{2\ell} (r_+\pm r_-)$, where $r_\pm$ are the values of the surface radius at outer/inner horizon \cite{Banados:1992gq,Banados:1992wn}. Generic soft hairy black hole solutions can be obtained from the BTZ black hole applying a generic ``soft boost'', i.e., acting on it with the full asymptotic symmetry group. Since soft boost generators commute with the Hamiltonian they do not change the energy.~\footnote{%
As explained below the softly boosted solution remains regular. This is in a stark contrast with what occurs if we apply the same procedure for 
Brown--Henneaux boundary conditions  \cite{Brown:1986nw}. This is because once acting on the solution with a generic Virasoro generator, not only the energy changes, but the regularity of the boosted solution is generically spoiled. Indeed, the additional Virasoro global charges of the boosted solution cannot be regarded as soft hair, because they do not commute with the Hamiltonian.
}

Soft hair charges do not contribute to the BH entropy, which can be readily computed from the Chern--Simons approach as in \cite{Perez:2012cf,Perez:2013xi,deBoer:2013gz,Bunster:2014mua}. In fact, the entropy of a generic soft hairy black hole is found to be given by
\eq{
S = 2\pi(J_{0}^{+}+J_{0}^{-}) 
 = \frac{A}{4G_N}
}{eq:Cardy}
This result naturally motivates performing a microstate counting in the spirit of \cite{Carlip:1994gy,Strominger:1997eq}. 

\section{Linking near horizon and asymptotic symmetries}

So far we have taken the perspective of a near horizon observer. Here we translate our findings into the language of an asymptotic observer. Since the global charges \eqref{eq:Qmn} and their algebra \eqref{eq:NHSA} do not depend on the radial coordinate, the same structure arises at infinity. We clarify below how our analysis manifests itself in terms of the standard variables in the asymptotic region.

Our near horizon boundary conditions \eqref{eq:r7}, \eqref{eq:r21} are written such that the auxiliary connections $\frak{a}^{\pm}$ are in diagonal gauge, while the standard asymptotic analysis uses the so-called highest weight gauge. Therefore, we transform the gauge fields in \eqref{eq:r7},~\eqref{eq:r21} to gauge fields $\hat{A}$ in the highest weight gauge.~\footnote{
To reduce clutter we discuss only one chiral part of the connections, $A^{+}$ and $\hat{A}^{+}$, and drop the superscript $+$.
}

For a generic choice of an unspecified chemical potential $\mu$ the
asymptotic form of the connection in the highest weight gauge is given
by \cite{Henneaux:2013dra,Bunster:2014mua}
\begin{align}
\hat{A} & =\hat{b}^{-1}\big(\extd+\hat{\frak{a}}\big)\hat{b}\qquad\hat{\frak{a}}_{\vp}=L_{1}-\tfrac{1}{2}\,{\cal L}\, L_{-1}\\
\hat{b} & =e^{\rho L_{0}}\quad\hat{\frak{a}}_{t}=\mu L_{1}-\mu^{\prime}L_{0}+\big(\tfrac{1}{2}\,\mu''-\tfrac{1}{2}\,{\cal L}\mu\big)\, L_{-1}\nonumber 
\end{align}
where ${\cal L}$ and $\mu$ are arbitrary functions of $t,\varphi$.

The problem reduces to find a  
gauge transformation
generated by a group element $g$, such that $\hat{\frak{a}}=g^{-1}\left(\extd+\frak{a}\right)g$,
followed by renaming the advanced time coordinate as $v=t$. We find $g=\exp{(xL_{1})}\cdot\exp{(-\tfrac{1}{2}{\cal J}L_{-1})}$,
where $x=x(v,\,\varphi)$ fulfills $\partial_vx-{\zeta}x=\mu$ and $x'-{\cal{J}}x=1$,
whose on-shell consistency implies
\begin{equation}
\mu^{\prime}-{\cal J}\mu=-\zeta\,.\label{eq:Mchp}
\end{equation}
Therefore, the asymptotic chemical potential $\mu$ depends not only on the near horizon chemical potential $\zeta$ but also on the near horizon charge ${\cal J}$.
The connections $\frak{a}$ and $\hat{\frak{a}}$ are mapped
to each other provided 
\begin{equation}
{\cal L}=\tfrac{1}{2}{\cal J}^{2}+{\cal J}^{\prime}\,.\label{eq:r36}
\end{equation}
We rephrase now the gravity result \eqref{eq:r36} algebraically.

If $\eta$ stands for
the parameter generating an arbitrary $\hat{u}(1)$ transformation,
$\delta{\cal {J}}=\eta^{\prime}$, then Eq.~\eqref{eq:Mchp} implies that
the corresponding parameter $\varepsilon$ in the highest weight gauge
depends on the global charges and fulfills $\varepsilon^{\prime}-{\cal J}\varepsilon=-\eta$.
Hence, according to \eqref{eq:r36}, the transformation law of ${\cal L}$ reads $\delta{\cal L}=2{\cal L}\varepsilon^{\prime}+{\cal L}^{\prime}\varepsilon-\varepsilon^{\prime\prime\prime}$.
Expanding in Fourier modes, Eq.~\eqref{eq:r36} yields
\begin{equation}
kL_{n}=\sum_{p\in\mathbb{Z}}J_{n-p}J_{p}+iknJ_{n}\,.\label{eq:rc43}
\end{equation}
This is a standard (twisted) Sugawara construction \cite{diFrancesco}.
The generators $L_{n}$ fulfill the Virasoro algebra with the Brown-Henneaux
central extension \cite{Brown:1986nw}
\begin{equation}
[L_{n},\, L_{m}]=(n-m)L_{n+m}+\tfrac{1}{2}\, k\, n^{3}\,\delta_{n+m,\,0}\;.\label{eq:Vir}
\end{equation}

What we have shown above is that the asymptotic symmetry algebra discovered
in \cite{Brown:1986nw} is composite from the near horizon perspective,
which can be interpreted as algebraic manifestation of black hole
complementarity \cite{'tHooft:1984re,Susskind:1993if,Stephens:1994an},
in the sense that the same physics is most naturally described in
very different terms for an asymptotic and a near horizon observer. 

Even though the spin-2 currents fulfill the Virasoro algebra \eqref{eq:Vir}, the corresponding global charges actually span the $\hat{u}(1)$ current algebra, which we show now explicitly. 
From the point of view of an observer at infinity, by
virtue of Eqs.~\eqref{eq:Mchp} and \eqref{eq:r36}, the variation
of the global charges reads 
\eq{
\delta{\cal Q}=-\frac{k}{4\pi}\oint \extd\varphi\,\varepsilon\,\delta\mathcal{L}=-\frac{k}{4\pi}\oint \extd\varphi\,\eta\,\delta\mathcal{J}\,.
}{eq:angelinajolie}
The global charges satisfy the near horizon symmetry algebra \eqref{eq:NHSA}. Thus, in spite
of the fact that the asymptotic conditions are written in the highest
weight gauge, the global charges do not fulfill the Virasoro algebra
with the Brown--Henneaux central extension, because the chemical potential
$\mu$ instead of being fixed at infinity without variation, for our
boundary conditions fulfill Eq.~\eqref{eq:Mchp}. In other words,
in our case $\mu$ explicitly depends on the global charges, while
what remains fixed at infinity is our chemical potential $\zeta$.

One remarkable feature of our boundary conditions is that they are
singled out by requiring that the whole spectrum is compatible with
regularity of the fields, regardless the value of the global charges.
Indeed, if the chemical potential $\zeta$ is assumed to be constant, and the topology of
the Euclidean manifold is that of a solid torus, where the contractible
cycle corresponds to Euclidean time, regularity of the gauge field
means that its holonomy around that cycle has to be trivial, 
implying
\begin{equation}
\mu\mu''-\tfrac{1}{2}\mu^{\prime\,2}-\mu^{2}{\cal L}=-2\pi^{2}/\beta^{2}\,,\label{eq:Reg}
\end{equation}
where $\beta$ is the length of the thermal cycle. 
The regularity condition \eqref{eq:Reg} is solved automatically
by virtue of the equations that define our boundary conditions,
\eqref{eq:Mchp} and \eqref{eq:r36}, provided $\zeta^{2}=4\pi^{2}/\beta^2$.
This last condition is easily obtained from solving the
regularity condition directly in the diagonal gauge \eqref{eq:r21} and for $\zeta^2=a^2$ amounts to the Unruh temperature $T=1/\beta=a/(2\pi)$ \cite{Unruh:1976db}.
Remarkably, no global charges are involved in this relationship.






\section{Concluding remarks}

One can use our near horizon algebra \eqref{eq:NHSA} to provide a microstate counting of the entropy \eqref{eq:Cardy} \cite{prep16}. Entropy can alternatively be calculated from composite algebras like the Virasoro algebra that we obtained in the last section. Indeed, in terms of the Virasoro zero modes $L_0^\pm$ using \eqref{eq:rc43} entropy \eqref{eq:Cardy} can be written in Cardy-form \cite{Cardy:1986ie,Bloete:1986qm} as $S=2\pi\sqrt{kL_{0}^{+}}+2\pi\sqrt{kL_{0}^{-}}$. Another way to perform a microstate counting through a composite algebra based on  our near horizon algebra \eqref{eq:NHSA} is to use the warped conformal algebra found in \cite{Donnay:2015abr} [their Eq.~(9)], which consists of a Virasoro and a $\hat{u}(1)$ current algebra. Introducing generators $J_n$ and $K_n$ as $J_n^{\pm}=\frac{1}{2}(J_n\pm{K}_n)$, the two algebras are related non-linearly through $Y_n \sim \,\sum_{p\in\mathbb{Z}} J_{n-p} K_p$ and $T_n = J_n$. Using known results pertaining to two-dimensional field theories invariant under a single Virasoro and a $\hat{u}(1)$ current algebra, called warped conformal field theories \cite{Detournay:2012pc}, the cylinder partition function written as $Z(\beta,\,\theta) = \mbox{Tr}\,e^{-{\beta}H+i{\theta}J}$, with $H=Q_{\partial_v}$ and $J=Q_{\partial_\varphi}$, enjoys the modular property
$Z(\beta,\,\theta)=Z(2\pi\beta/\theta,\,-4\pi^2/\theta)$,
which allows to project the partition function on the ground state at small imaginary $\theta$  \cite{Detournay:2012pc}, yielding
an entropy
$S=2\pi\beta\,H^{\textrm{\tiny{vac}}}/\theta+i8\pi^2\,J^{\textrm{\tiny{vac}}}/\theta$.
Assuming the vacuum state has no angular momentum, $J^{\textrm{\tiny{vac}}}=0$, and using $H=-\partial\ln{Z}/\partial\beta=2{\pi}H^{\textrm{\tiny{vac}}}/\theta$  
establishes
\eq{
S = \beta H = 
\frac{A}{4G_N} = S_{\textrm{\tiny BH}}\,.
}{FT10}
Interestingly, this result is independent from $H^{\textrm{\tiny{vac}}}$. With $\beta=2\pi/a$ and $H=aP_0=\frac{1}{8{\pi}G}\oint\extd\varphi\,\gamma=\frac{A}{8{\pi}G}$, we recover the BH entropy law \eqref{eq:r1}. This provides a microscopic explanation for the observation \cite{Donnay:2015abr} that $H$ is the product of black hole entropy and temperature.

Our results easily extend to the case
of general relativity in 3d without cosmological constant ($\ell\to\infty$)
\cite{prep16}. In particular, the flat limit of the metric (\ref{eq:r5})
describes an interesting class of ``soft hairy cosmological
spacetimes'' that contains the solutions discussed in \cite{Cornalba:2002fi}.
Also in that case we find that soft hair charges do not contribute to the
entropy, which in turn agrees with the asymptotic state counting in \cite{Barnich:2012xq,Bagchi:2012xr}.
The asymptotically flat structure can be recovered in the limit along
the lines of \cite{Barnich:2013yka,Afshar:2013bla,Gary:2014ppa,Matulich:2014hea}. We find the near horizon symmetry algebra 
\begin{equation}
\left[J_{n},J_{m}\right]=\left[K_{n},K_{m}\right]=0\quad\left[J_{n},K_{m}\right]=kn\delta_{n+m,0}\,.\label{eq:ads1}
\end{equation}
The Hamiltonian again commutes with all the generators, and therefore
with all descendants of the vacuum. The centrally-extended BMS
currents \cite{Ashtekar:1996cd,Barnich:2006av} are recovered
as composite operators constructed out from \eqref{eq:ads1}. 

It is clear that the specifics of our construction, including soft
hair and black hole complementarity, certainly apply to different
3d gravity theories whose field equations are solved for constant
curvature spacetimes, as it is the case of the massive gravity theories
discussed in \cite{Deser:1982vy,Bergshoeff:2009hq,Deser:2009hb,Bergshoeff:2013xma},
conformal gravity \cite{Afshar:2011yh,Afshar:2011qw,Afshar:2013bla} and generalizations thereof, as well as for higher
spin gravity in AdS \cite{Henneaux:2010xg,Campoleoni:2010zq,Gaberdiel:2011wb,Henneaux:2013dra,Bunster:2014mua}
or in flat space \cite{Afshar:2013vka,Gonzalez:2013oaa,Gary:2014ppa,Matulich:2014hea}. 
It will be interesting to recover these physical features in four dimensions \cite{Bondi:1962,Sachs:1962,Barnich:2009se,Donnay:2015abr,Hawking:2016msc}.


\section*{Acknowledgments}

\acknowledgments

We are grateful to Gaston Giribet, Hern\'an
Gonz\'alez, Javier Matulich, Miguel Pino, Stefan Prohazka, Max Riegler, Jakob Salzer,
Friedrich Sch\"oller and C\'edric Troessaert for discussions.

HA was supported in part by the Boniad-Melli-Nokhebgan (Iran's National Elite Foundation, INEF).
SD is a Research Associate of the Fonds de la Recherche Scientifique
F.R.S.-FNRS (Belgium). He is supported in part by the ARC grant ``Holography,
Gauge Theories and Quantum Gravity Building models of quantum black
holes'', by IISN - Belgium (convention 4.4503.15) and benefited from
the support of the Solvay Family. DG was supported by the Austrian
Science Fund (FWF), project P~27182-N27 and by CECS. WM was supported
by the FWF project P~27182-N27. The work of AP, DT and RT is
partially funded by the Fondecyt grants Nr.~11130262, 11130260, 1130658,
1121031. The Centro de Estudios Cient\'ificos (CECs) is funded by the
Chilean Government through the Centers of Excellence Base Financing
Program of Conicyt.


\begin{thebibliography}{64}%
\makeatletter
\providecommand \@ifxundefined [1]{%
 \@ifx{#1\undefined}
}%
\providecommand \@ifnum [1]{%
 \ifnum #1\expandafter \@firstoftwo
 \else \expandafter \@secondoftwo
 \fi
}%
\providecommand \@ifx [1]{%
 \ifx #1\expandafter \@firstoftwo
 \else \expandafter \@secondoftwo
 \fi
}%
\providecommand \natexlab [1]{#1}%
\providecommand \enquote  [1]{``#1''}%
\providecommand \bibnamefont  [1]{#1}%
\providecommand \bibfnamefont [1]{#1}%
\providecommand \citenamefont [1]{#1}%
\providecommand \href@noop [0]{\@secondoftwo}%
\providecommand \href [0]{\begingroup \@sanitize@url \@href}%
\providecommand \@href[1]{\@@startlink{#1}\@@href}%
\providecommand \@@href[1]{\endgroup#1\@@endlink}%
\providecommand \@sanitize@url [0]{\catcode `\\12\catcode `\$12\catcode
  `\&12\catcode `\#12\catcode `\^12\catcode `\_12\catcode `\%12\relax}%
\providecommand \@@startlink[1]{}%
\providecommand \@@endlink[0]{}%
\providecommand \url  [0]{\begingroup\@sanitize@url \@url }%
\providecommand \@url [1]{\endgroup\@href {#1}{\urlprefix }}%
\providecommand \urlprefix  [0]{URL }%
\providecommand \Eprint [0]{\href }%
\providecommand \doibase [0]{http://dx.doi.org/}%
\providecommand \selectlanguage [0]{\@gobble}%
\providecommand \bibinfo  [0]{\@secondoftwo}%
\providecommand \bibfield  [0]{\@secondoftwo}%
\providecommand \translation [1]{[#1]}%
\providecommand \BibitemOpen [0]{}%
\providecommand \bibitemStop [0]{}%
\providecommand \bibitemNoStop [0]{.\EOS\space}%
\providecommand \EOS [0]{\spacefactor3000\relax}%
\providecommand \BibitemShut  [1]{\csname bibitem#1\endcsname}%
\let\auto@bib@innerbib\@empty
\bibitem [{\citenamefont {Carlip}(1995)}]{Carlip:1994gy}%
  \BibitemOpen
  \bibfield  {author} {\bibinfo {author} {\bibfnamefont {S.}~\bibnamefont
  {Carlip}},\ }\href@noop {} {\bibfield  {journal} {\bibinfo  {journal} {Phys.
  Rev.}\ }\textbf {\bibinfo {volume} {D51}},\ \bibinfo {pages} {632} (\bibinfo
  {year} {1995})},\ \Eprint {http://arxiv.org/abs/gr-qc/9409052}
  {gr-qc/9409052} \BibitemShut {NoStop}%
\bibitem [{\citenamefont {Strominger}\ and\ \citenamefont
  {Vafa}(1996)}]{Strominger:1996sh}%
  \BibitemOpen
  \bibfield  {author} {\bibinfo {author} {\bibfnamefont {A.}~\bibnamefont
  {Strominger}}\ and\ \bibinfo {author} {\bibfnamefont {C.}~\bibnamefont
  {Vafa}},\ }\href@noop {} {\bibfield  {journal} {\bibinfo  {journal} {Phys.
  Lett.}\ }\textbf {\bibinfo {volume} {B379}},\ \bibinfo {pages} {99} (\bibinfo
  {year} {1996})},\ \Eprint {http://arxiv.org/abs/hep-th/9601029}
  {hep-th/9601029} \BibitemShut {NoStop}%
\bibitem [{\citenamefont {Strominger}(1998)}]{Strominger:1997eq}%
  \BibitemOpen
  \bibfield  {author} {\bibinfo {author} {\bibfnamefont {A.}~\bibnamefont
  {Strominger}},\ }\href@noop {} {\bibfield  {journal} {\bibinfo  {journal}
  {JHEP}\ }\textbf {\bibinfo {volume} {02}},\ \bibinfo {pages} {009} (\bibinfo
  {year} {1998})},\ \Eprint {http://arxiv.org/abs/hep-th/9712251}
  {hep-th/9712251} \BibitemShut {NoStop}%
\bibitem [{\citenamefont {Carlip}(1999)}]{Carlip:1998wz}%
  \BibitemOpen
  \bibfield  {author} {\bibinfo {author} {\bibfnamefont {S.}~\bibnamefont
  {Carlip}},\ }\href@noop {} {\bibfield  {journal} {\bibinfo  {journal} {Phys.
  Rev. Lett.}\ }\textbf {\bibinfo {volume} {82}},\ \bibinfo {pages} {2828}
  (\bibinfo {year} {1999})},\ \Eprint {http://arxiv.org/abs/hep-th/9812013}
  {hep-th/9812013} \BibitemShut {NoStop}%
\bibitem [{\citenamefont {Guica}\ \emph {et~al.}(2009)\citenamefont {Guica},
  \citenamefont {Hartman}, \citenamefont {Song},\ and\ \citenamefont
  {Strominger}}]{Guica:2008mu}%
  \BibitemOpen
  \bibfield  {author} {\bibinfo {author} {\bibfnamefont {M.}~\bibnamefont
  {Guica}}, \bibinfo {author} {\bibfnamefont {T.}~\bibnamefont {Hartman}},
  \bibinfo {author} {\bibfnamefont {W.}~\bibnamefont {Song}}, \ and\ \bibinfo
  {author} {\bibfnamefont {A.}~\bibnamefont {Strominger}},\ }\href {\doibase
  10.1103/PhysRevD.80.124008} {\bibfield  {journal} {\bibinfo  {journal} {Phys.
  Rev.}\ }\textbf {\bibinfo {volume} {D80}},\ \bibinfo {pages} {124008}
  (\bibinfo {year} {2009})},\ \Eprint {http://arxiv.org/abs/0809.4266}
  {arXiv:0809.4266 [hep-th]} \BibitemShut {NoStop}%
\bibitem [{\citenamefont {Bondi}\ \emph {et~al.}(1962)\citenamefont {Bondi},
  \citenamefont {van~der Burg},\ and\ \citenamefont {Metzner}}]{Bondi:1962}%
  \BibitemOpen
  \bibfield  {author} {\bibinfo {author} {\bibfnamefont {H.}~\bibnamefont
  {Bondi}}, \bibinfo {author} {\bibfnamefont {M.}~\bibnamefont {van~der Burg}},
  \ and\ \bibinfo {author} {\bibfnamefont {A.}~\bibnamefont {Metzner}},\
  }\href@noop {} {\bibfield  {journal} {\bibinfo  {journal} {Proc. Roy. Soc.
  London}\ }\textbf {\bibinfo {volume} {A269}},\ \bibinfo {pages} {21}
  (\bibinfo {year} {1962})}\BibitemShut {NoStop}%
\bibitem [{\citenamefont {Sachs}(1962)}]{Sachs:1962}%
  \BibitemOpen
  \bibfield  {author} {\bibinfo {author} {\bibfnamefont {R.}~\bibnamefont
  {Sachs}},\ }\href@noop {} {\bibfield  {journal} {\bibinfo  {journal} {Phys.
  Rev.}\ }\textbf {\bibinfo {volume} {128}},\ \bibinfo {pages} {2851} (\bibinfo
  {year} {1962})}\BibitemShut {NoStop}%
\bibitem [{\citenamefont {Ashtekar}\ \emph {et~al.}(1997)\citenamefont
  {Ashtekar}, \citenamefont {Bicak},\ and\ \citenamefont
  {Schmidt}}]{Ashtekar:1996cd}%
  \BibitemOpen
  \bibfield  {author} {\bibinfo {author} {\bibfnamefont {A.}~\bibnamefont
  {Ashtekar}}, \bibinfo {author} {\bibfnamefont {J.}~\bibnamefont {Bicak}}, \
  and\ \bibinfo {author} {\bibfnamefont {B.~G.}\ \bibnamefont {Schmidt}},\
  }\href {\doibase 10.1103/PhysRevD.55.669} {\bibfield  {journal} {\bibinfo
  {journal} {Phys.Rev.}\ }\textbf {\bibinfo {volume} {D55}},\ \bibinfo {pages}
  {669} (\bibinfo {year} {1997})},\ \Eprint
  {http://arxiv.org/abs/gr-qc/9608042} {arXiv:gr-qc/9608042 [gr-qc]}
  \BibitemShut {NoStop}%
\bibitem [{\citenamefont {Barnich}\ and\ \citenamefont
  {Comp{\`e}re}(2007)}]{Barnich:2006av}%
  \BibitemOpen
  \bibfield  {author} {\bibinfo {author} {\bibfnamefont {G.}~\bibnamefont
  {Barnich}}\ and\ \bibinfo {author} {\bibfnamefont {G.}~\bibnamefont
  {Comp{\`e}re}},\ }\href {\doibase 10.1088/0264-9381/24/5/F01,
  10.1088/0264-9381/24/11/C01} {\bibfield  {journal} {\bibinfo  {journal}
  {Class.Quant.Grav.}\ }\textbf {\bibinfo {volume} {24}},\ \bibinfo {pages}
  {F15} (\bibinfo {year} {2007})},\ \Eprint
  {http://arxiv.org/abs/gr-qc/0610130} {arXiv:gr-qc/0610130 [gr-qc]}
  \BibitemShut {NoStop}%
\bibitem [{\citenamefont {Donnay}\ \emph {et~al.}(2016)\citenamefont {Donnay},
  \citenamefont {Giribet}, \citenamefont {Gonzalez},\ and\ \citenamefont
  {Pino}}]{Donnay:2015abr}%
  \BibitemOpen
  \bibfield  {author} {\bibinfo {author} {\bibfnamefont {L.}~\bibnamefont
  {Donnay}}, \bibinfo {author} {\bibfnamefont {G.}~\bibnamefont {Giribet}},
  \bibinfo {author} {\bibfnamefont {H.~A.}\ \bibnamefont {Gonzalez}}, \ and\
  \bibinfo {author} {\bibfnamefont {M.}~\bibnamefont {Pino}},\ }\href {\doibase
  10.1103/PhysRevLett.116.091101} {\bibfield  {journal} {\bibinfo  {journal}
  {Phys. Rev. Lett.}\ }\textbf {\bibinfo {volume} {116}},\ \bibinfo {pages}
  {091101} (\bibinfo {year} {2016})},\ \Eprint
  {http://arxiv.org/abs/1511.08687} {arXiv:1511.08687 [hep-th]} \BibitemShut
  {NoStop}%
\bibitem [{\citenamefont {Afshar}\ \emph {et~al.}(2015)\citenamefont {Afshar},
  \citenamefont {Detournay}, \citenamefont {Grumiller},\ and\ \citenamefont
  {Oblak}}]{Afshar:2015wjm}%
  \BibitemOpen
  \bibfield  {author} {\bibinfo {author} {\bibfnamefont {H.}~\bibnamefont
  {Afshar}}, \bibinfo {author} {\bibfnamefont {S.}~\bibnamefont {Detournay}},
  \bibinfo {author} {\bibfnamefont {D.}~\bibnamefont {Grumiller}}, \ and\
  \bibinfo {author} {\bibfnamefont {B.}~\bibnamefont {Oblak}},\ }\href@noop {}
  {\  (\bibinfo {year} {2015})},\ \Eprint {http://arxiv.org/abs/1512.08233}
  {arXiv:1512.08233 [hep-th]} \BibitemShut {NoStop}%
\bibitem [{\citenamefont {Hawking}\ \emph {et~al.}(2016)\citenamefont
  {Hawking}, \citenamefont {Perry},\ and\ \citenamefont
  {Strominger}}]{Hawking:2016msc}%
  \BibitemOpen
  \bibfield  {author} {\bibinfo {author} {\bibfnamefont {S.~W.}\ \bibnamefont
  {Hawking}}, \bibinfo {author} {\bibfnamefont {M.~J.}\ \bibnamefont {Perry}},
  \ and\ \bibinfo {author} {\bibfnamefont {A.}~\bibnamefont {Strominger}},\
  }\href@noop {} {\  (\bibinfo {year} {2016})},\ \Eprint
  {http://arxiv.org/abs/1601.00921} {arXiv:1601.00921 [hep-th]} \BibitemShut
  {NoStop}%
\bibitem [{\citenamefont {Blau}\ and\ \citenamefont
  {O'Loughlin}(2015)}]{Blau:2015nee}%
  \BibitemOpen
  \bibfield  {author} {\bibinfo {author} {\bibfnamefont {M.}~\bibnamefont
  {Blau}}\ and\ \bibinfo {author} {\bibfnamefont {M.}~\bibnamefont
  {O'Loughlin}},\ }\href@noop {} {\  (\bibinfo {year} {2015})},\ \Eprint
  {http://arxiv.org/abs/1512.02858} {arXiv:1512.02858 [hep-th]} \BibitemShut
  {NoStop}%
\bibitem [{\citenamefont {Penna}(2015)}]{Penna:2015gza}%
  \BibitemOpen
  \bibfield  {author} {\bibinfo {author} {\bibfnamefont {R.~F.}\ \bibnamefont
  {Penna}},\ }\href@noop {} {\  (\bibinfo {year} {2015})},\ \Eprint
  {http://arxiv.org/abs/1508.06577} {arXiv:1508.06577 [hep-th]} \BibitemShut
  {NoStop}%
\bibitem [{\citenamefont {Hooft}(2016)}]{Hooft:2016itl}%
  \BibitemOpen
  \bibfield  {author} {\bibinfo {author} {\bibfnamefont {G.~t.}\ \bibnamefont
  {Hooft}},\ }\href@noop {} {\  (\bibinfo {year} {2016})},\ \Eprint
  {http://arxiv.org/abs/1601.03447} {arXiv:1601.03447 [gr-qc]} \BibitemShut
  {NoStop}%
\bibitem [{\citenamefont {Bianchi}\ and\ \citenamefont
  {Guerrieri}(2016)}]{Bianchi:2016tju}%
  \BibitemOpen
  \bibfield  {author} {\bibinfo {author} {\bibfnamefont {M.}~\bibnamefont
  {Bianchi}}\ and\ \bibinfo {author} {\bibfnamefont {A.~L.}\ \bibnamefont
  {Guerrieri}},\ }in\ \href
  {https://inspirehep.net/record/1415295/files/arXiv:1601.03457.pdf} {\emph
  {\bibinfo {booktitle} {{14th Marcel Grossmann Meeting 
  2015}}}}\
  (\bibinfo {year} {2016})\ \Eprint {http://arxiv.org/abs/1601.03457}
  {arXiv:1601.03457 [hep-th]} \BibitemShut {NoStop}%
\bibitem [{\citenamefont {Averin}\ \emph {et~al.}(2016)\citenamefont {Averin},
  \citenamefont {Dvali}, \citenamefont {Gomez},\ and\ \citenamefont
  {Lust}}]{Averin:2016ybl}%
  \BibitemOpen
  \bibfield  {author} {\bibinfo {author} {\bibfnamefont {A.}~\bibnamefont
  {Averin}}, \bibinfo {author} {\bibfnamefont {G.}~\bibnamefont {Dvali}},
  \bibinfo {author} {\bibfnamefont {C.}~\bibnamefont {Gomez}}, \ and\ \bibinfo
  {author} {\bibfnamefont {D.}~\bibnamefont {Lust}},\ }\href@noop {} {\
  (\bibinfo {year} {2016})},\ \Eprint {http://arxiv.org/abs/1601.03725}
  {arXiv:1601.03725 [hep-th]} \BibitemShut {NoStop}%
\bibitem [{\citenamefont {Comp{\`e}re}\ and\ \citenamefont
  {Long}(2016)}]{Compere:2016jwb}%
  \BibitemOpen
  \bibfield  {author} {\bibinfo {author} {\bibfnamefont {G.}~\bibnamefont
  {Comp{\`e}re}}\ and\ \bibinfo {author} {\bibfnamefont {J.}~\bibnamefont
  {Long}},\ }\href@noop {} {\  (\bibinfo {year} {2016})},\ \Eprint
  {http://arxiv.org/abs/1601.04958} {arXiv:1601.04958 [hep-th]} \BibitemShut
  {NoStop}%
\bibitem [{\citenamefont {Kehagias}\ and\ \citenamefont
  {Riotto}(2016)}]{Kehagias:2016zry}%
  \BibitemOpen
  \bibfield  {author} {\bibinfo {author} {\bibfnamefont {A.}~\bibnamefont
  {Kehagias}}\ and\ \bibinfo {author} {\bibfnamefont {A.}~\bibnamefont
  {Riotto}},\ }\href@noop {} {\  (\bibinfo {year} {2016})},\ \Eprint
  {http://arxiv.org/abs/1602.02653} {arXiv:1602.02653 [hep-th]} \BibitemShut
  {NoStop}%
\bibitem [{\citenamefont {Comp\`ere}\ and\ \citenamefont
  {Long}(2016)}]{Compere:2016hzt}%
  \BibitemOpen
  \bibfield  {author} {\bibinfo {author} {\bibfnamefont {G.}~\bibnamefont
  {Comp\`ere}}\ and\ \bibinfo {author} {\bibfnamefont {J.}~\bibnamefont
  {Long}},\ }\href@noop {} {\  (\bibinfo {year} {2016})},\ \Eprint
  {http://arxiv.org/abs/1602.05197} {arXiv:1602.05197 [gr-qc]} \BibitemShut
  {NoStop}%
\bibitem [{\citenamefont {Rindler}(1966)}]{Rindler:1966zz}%
  \BibitemOpen
  \bibfield  {author} {\bibinfo {author} {\bibfnamefont {W.}~\bibnamefont
  {Rindler}},\ }\href {\doibase 10.1119/1.1972547} {\bibfield  {journal}
  {\bibinfo  {journal} {Am.J.Phys.}\ }\textbf {\bibinfo {volume} {34}},\
  \bibinfo {pages} {1174} (\bibinfo {year} {1966})}\BibitemShut {NoStop}%
\bibitem [{\citenamefont {Brown}\ and\ \citenamefont
  {Henneaux}(1986)}]{Brown:1986nw}%
  \BibitemOpen
  \bibfield  {author} {\bibinfo {author} {\bibfnamefont {J.~D.}\ \bibnamefont
  {Brown}}\ and\ \bibinfo {author} {\bibfnamefont {M.}~\bibnamefont
  {Henneaux}},\ }\href@noop {} {\bibfield  {journal} {\bibinfo  {journal}
  {Commun. Math. Phys.}\ }\textbf {\bibinfo {volume} {104}},\ \bibinfo {pages}
  {207} (\bibinfo {year} {1986})}\BibitemShut {NoStop}%
\bibitem [{\citenamefont {'t~Hooft}(1985)}]{'tHooft:1984re}%
  \BibitemOpen
  \bibfield  {author} {\bibinfo {author} {\bibfnamefont {G.}~\bibnamefont
  {'t~Hooft}},\ }\href {\doibase 10.1016/0550-3213(85)90418-3} {\bibfield
  {journal} {\bibinfo  {journal} {Nucl. Phys.}\ }\textbf {\bibinfo {volume}
  {B256}},\ \bibinfo {pages} {727} (\bibinfo {year} {1985})}\BibitemShut
  {NoStop}%
\bibitem [{\citenamefont {Susskind}\ \emph {et~al.}(1993)\citenamefont
  {Susskind}, \citenamefont {Thorlacius},\ and\ \citenamefont
  {Uglum}}]{Susskind:1993if}%
  \BibitemOpen
  \bibfield  {author} {\bibinfo {author} {\bibfnamefont {L.}~\bibnamefont
  {Susskind}}, \bibinfo {author} {\bibfnamefont {L.}~\bibnamefont
  {Thorlacius}}, \ and\ \bibinfo {author} {\bibfnamefont {J.}~\bibnamefont
  {Uglum}},\ }\href {\doibase 10.1103/PhysRevD.48.3743} {\bibfield  {journal}
  {\bibinfo  {journal} {Phys.Rev.}\ }\textbf {\bibinfo {volume} {D48}},\
  \bibinfo {pages} {3743} (\bibinfo {year} {1993})},\ \Eprint
  {http://arxiv.org/abs/hep-th/9306069} {arXiv:hep-th/9306069 [hep-th]}
  \BibitemShut {NoStop}%
\bibitem [{\citenamefont {Stephens}\ \emph {et~al.}(1994)\citenamefont
  {Stephens}, \citenamefont {'t~Hooft},\ and\ \citenamefont
  {Whiting}}]{Stephens:1994an}%
  \BibitemOpen
  \bibfield  {author} {\bibinfo {author} {\bibfnamefont {C.~R.}\ \bibnamefont
  {Stephens}}, \bibinfo {author} {\bibfnamefont {G.}~\bibnamefont {'t~Hooft}},
  \ and\ \bibinfo {author} {\bibfnamefont {B.~F.}\ \bibnamefont {Whiting}},\
  }\href@noop {} {\bibfield  {journal} {\bibinfo  {journal} {Class. Quant.
  Grav.}\ }\textbf {\bibinfo {volume} {11}},\ \bibinfo {pages} {621} (\bibinfo
  {year} {1994})},\ \Eprint {http://arXiv.org/abs/gr-qc/9310006}
  {gr-qc/9310006} \BibitemShut {NoStop}%
\bibitem [{\citenamefont {Barnich}\ \emph {et~al.}(2015)\citenamefont
  {Barnich}, \citenamefont {Troessaert}, \citenamefont {Tempo},\ and\
  \citenamefont {Troncoso}}]{Barnich:2015dvt}%
  \BibitemOpen
  \bibfield  {author} {\bibinfo {author} {\bibfnamefont {G.}~\bibnamefont
  {Barnich}}, \bibinfo {author} {\bibfnamefont {C.}~\bibnamefont {Troessaert}},
  \bibinfo {author} {\bibfnamefont {D.}~\bibnamefont {Tempo}}, \ and\ \bibinfo
  {author} {\bibfnamefont {R.}~\bibnamefont {Troncoso}},\ }\href@noop {} {\
  (\bibinfo {year} {2015})},\ \Eprint {http://arxiv.org/abs/1512.05410}
  {arXiv:1512.05410 [hep-th]} \BibitemShut {NoStop}%
\bibitem [{\citenamefont {Ba\~nados}\ \emph {et~al.}(1992)\citenamefont
  {Ba\~nados}, \citenamefont {Teitelboim},\ and\ \citenamefont
  {Zanelli}}]{Banados:1992wn}%
  \BibitemOpen
  \bibfield  {author} {\bibinfo {author} {\bibfnamefont {M.}~\bibnamefont
  {Ba\~nados}}, \bibinfo {author} {\bibfnamefont {C.}~\bibnamefont
  {Teitelboim}}, \ and\ \bibinfo {author} {\bibfnamefont {J.}~\bibnamefont
  {Zanelli}},\ }\href@noop {} {\bibfield  {journal} {\bibinfo  {journal} {Phys.
  Rev. Lett.}\ }\textbf {\bibinfo {volume} {69}},\ \bibinfo {pages} {1849}
  (\bibinfo {year} {1992})},\ \Eprint {http://arxiv.org/abs/hep-th/9204099}
  {hep-th/9204099} \BibitemShut {NoStop}%
\bibitem [{\citenamefont {Ba\~nados}\ \emph {et~al.}(1993)\citenamefont
  {Ba\~nados}, \citenamefont {Henneaux}, \citenamefont {Teitelboim},\ and\
  \citenamefont {Zanelli}}]{Banados:1992gq}%
  \BibitemOpen
  \bibfield  {author} {\bibinfo {author} {\bibfnamefont {M.}~\bibnamefont
  {Ba\~nados}}, \bibinfo {author} {\bibfnamefont {M.}~\bibnamefont {Henneaux}},
  \bibinfo {author} {\bibfnamefont {C.}~\bibnamefont {Teitelboim}}, \ and\
  \bibinfo {author} {\bibfnamefont {J.}~\bibnamefont {Zanelli}},\ }\href@noop
  {} {\bibfield  {journal} {\bibinfo  {journal} {Phys. Rev.}\ }\textbf
  {\bibinfo {volume} {D48}},\ \bibinfo {pages} {1506} (\bibinfo {year}
  {1993})},\ \Eprint {http://arxiv.org/abs/gr-qc/9302012} {gr-qc/9302012}
  \BibitemShut {NoStop}%
\bibitem [{\citenamefont {Achucarro}\ and\ \citenamefont
  {Townsend}(1986)}]{Achucarro:1987vz}%
  \BibitemOpen
  \bibfield  {author} {\bibinfo {author} {\bibfnamefont {A.}~\bibnamefont
  {Achucarro}}\ and\ \bibinfo {author} {\bibfnamefont {P.~K.}\ \bibnamefont
  {Townsend}},\ }\href {\doibase 10.1016/0370-2693(86)90140-1} {\bibfield
  {journal} {\bibinfo  {journal} {Phys. Lett.}\ }\textbf {\bibinfo {volume}
  {B180}},\ \bibinfo {pages} {89} (\bibinfo {year} {1986})}\BibitemShut
  {NoStop}%
\bibitem [{\citenamefont {Witten}(1988)}]{Witten:1988hc}%
  \BibitemOpen
  \bibfield  {author} {\bibinfo {author} {\bibfnamefont {E.}~\bibnamefont
  {Witten}},\ }\href {\doibase 10.1016/0550-3213(88)90143-5} {\bibfield
  {journal} {\bibinfo  {journal} {Nucl. Phys.}\ }\textbf {\bibinfo {volume}
  {B311}},\ \bibinfo {pages} {46} (\bibinfo {year} {1988})}\BibitemShut
  {NoStop}%
\bibitem [{\citenamefont {Comp{\`e}re}\ \emph {et~al.}(2013)\citenamefont
  {Comp{\`e}re}, \citenamefont {Jottar},\ and\ \citenamefont
  {Song}}]{Compere:2013nba}%
  \BibitemOpen
  \bibfield  {author} {\bibinfo {author} {\bibfnamefont {G.}~\bibnamefont
  {Comp{\`e}re}}, \bibinfo {author} {\bibfnamefont {J.~I.}\ \bibnamefont
  {Jottar}}, \ and\ \bibinfo {author} {\bibfnamefont {W.}~\bibnamefont
  {Song}},\ }\href {\doibase 10.1007/JHEP11(2013)054} {\bibfield  {journal}
  {\bibinfo  {journal} {JHEP}\ }\textbf {\bibinfo {volume} {1311}},\ \bibinfo
  {pages} {054} (\bibinfo {year} {2013})},\ \Eprint
  {http://arxiv.org/abs/1308.2175} {arXiv:1308.2175 [hep-th]} \BibitemShut
  {NoStop}%
\bibitem [{\citenamefont {Henneaux}\ \emph {et~al.}(2013)\citenamefont
  {Henneaux}, \citenamefont {Perez}, \citenamefont {Tempo},\ and\ \citenamefont
  {Troncoso}}]{Henneaux:2013dra}%
  \BibitemOpen
  \bibfield  {author} {\bibinfo {author} {\bibfnamefont {M.}~\bibnamefont
  {Henneaux}}, \bibinfo {author} {\bibfnamefont {A.}~\bibnamefont {Perez}},
  \bibinfo {author} {\bibfnamefont {D.}~\bibnamefont {Tempo}}, \ and\ \bibinfo
  {author} {\bibfnamefont {R.}~\bibnamefont {Troncoso}},\ }\href {\doibase
  10.1007/JHEP12(2013)048} {\bibfield  {journal} {\bibinfo  {journal} {JHEP}\
  }\textbf {\bibinfo {volume} {1312}},\ \bibinfo {pages} {048} (\bibinfo {year}
  {2013})},\ \Eprint {http://arxiv.org/abs/1309.4362} {arXiv:1309.4362
  [hep-th]} \BibitemShut {NoStop}%
\bibitem [{\citenamefont {Regge}\ and\ \citenamefont
  {Teitelboim}(1974)}]{Regge:1974zd}%
  \BibitemOpen
  \bibfield  {author} {\bibinfo {author} {\bibfnamefont {T.}~\bibnamefont
  {Regge}}\ and\ \bibinfo {author} {\bibfnamefont {C.}~\bibnamefont
  {Teitelboim}},\ }\href@noop {} {\bibfield  {journal} {\bibinfo  {journal}
  {Ann. Phys.}\ }\textbf {\bibinfo {volume} {88}},\ \bibinfo {pages} {286}
  (\bibinfo {year} {1974})}\BibitemShut {NoStop}%
\bibitem [{\citenamefont {Ba\~nados}(1995)}]{Banados:1994tn}%
  \BibitemOpen
  \bibfield  {author} {\bibinfo {author} {\bibfnamefont {M.}~\bibnamefont
  {Ba\~nados}},\ }\href@noop {} {\bibfield  {journal} {\bibinfo  {journal}
  {Phys. Rev.}\ }\textbf {\bibinfo {volume} {D52}},\ \bibinfo {pages} {5816}
  (\bibinfo {year} {1995})},\ \Eprint {http://arxiv.org/abs/hep-th/9405171}
  {hep-th/9405171} \BibitemShut {NoStop}%
\bibitem [{\citenamefont {Benguria}\ \emph {et~al.}(1977)\citenamefont
  {Benguria}, \citenamefont {Cordero},\ and\ \citenamefont
  {Teitelboim}}]{Benguria:1976in}%
  \BibitemOpen
  \bibfield  {author} {\bibinfo {author} {\bibfnamefont {R.}~\bibnamefont
  {Benguria}}, \bibinfo {author} {\bibfnamefont {P.}~\bibnamefont {Cordero}}, \
  and\ \bibinfo {author} {\bibfnamefont {C.}~\bibnamefont {Teitelboim}},\
  }\href {\doibase 10.1016/0550-3213(77)90426-6} {\bibfield  {journal}
  {\bibinfo  {journal} {Nucl. Phys.}\ }\textbf {\bibinfo {volume} {B122}},\
  \bibinfo {pages} {61} (\bibinfo {year} {1977})}\BibitemShut {NoStop}%
\bibitem [{\citenamefont {Perez}\ \emph
  {et~al.}(2013{\natexlab{a}})\citenamefont {Perez}, \citenamefont {Tempo},\
  and\ \citenamefont {Troncoso}}]{Perez:2012cf}%
  \BibitemOpen
  \bibfield  {author} {\bibinfo {author} {\bibfnamefont {A.}~\bibnamefont
  {Perez}}, \bibinfo {author} {\bibfnamefont {D.}~\bibnamefont {Tempo}}, \ and\
  \bibinfo {author} {\bibfnamefont {R.}~\bibnamefont {Troncoso}},\ }\href
  {\doibase 10.1016/j.physletb.2013.08.038} {\bibfield  {journal} {\bibinfo
  {journal} {Phys. Lett.}\ }\textbf {\bibinfo {volume} {B726}},\ \bibinfo
  {pages} {444} (\bibinfo {year} {2013}{\natexlab{a}})},\ \Eprint
  {http://arxiv.org/abs/1207.2844} {arXiv:1207.2844 [hep-th]} \BibitemShut
  {NoStop}%
\bibitem [{\citenamefont {Perez}\ \emph
  {et~al.}(2013{\natexlab{b}})\citenamefont {Perez}, \citenamefont {Tempo},\
  and\ \citenamefont {Troncoso}}]{Perez:2013xi}%
  \BibitemOpen
  \bibfield  {author} {\bibinfo {author} {\bibfnamefont {A.}~\bibnamefont
  {Perez}}, \bibinfo {author} {\bibfnamefont {D.}~\bibnamefont {Tempo}}, \ and\
  \bibinfo {author} {\bibfnamefont {R.}~\bibnamefont {Troncoso}},\ }\href@noop
  {} {\  (\bibinfo {year} {2013}{\natexlab{b}})},\ \Eprint
  {http://arxiv.org/abs/1301.0847} {arXiv:1301.0847 [hep-th]} \BibitemShut
  {NoStop}%
\bibitem [{\citenamefont {de~Boer}\ and\ \citenamefont
  {Jottar}(2014)}]{deBoer:2013gz}%
  \BibitemOpen
  \bibfield  {author} {\bibinfo {author} {\bibfnamefont {J.}~\bibnamefont
  {de~Boer}}\ and\ \bibinfo {author} {\bibfnamefont {J.~I.}\ \bibnamefont
  {Jottar}},\ }\href {\doibase 10.1007/JHEP01(2014)023} {\bibfield  {journal}
  {\bibinfo  {journal} {JHEP}\ }\textbf {\bibinfo {volume} {1401}},\ \bibinfo
  {pages} {023} (\bibinfo {year} {2014})},\ \Eprint
  {http://arxiv.org/abs/1302.0816} {arXiv:1302.0816 [hep-th]} \BibitemShut
  {NoStop}%
\bibitem [{\citenamefont {Bunster}\ \emph {et~al.}(2014)\citenamefont
  {Bunster}, \citenamefont {Henneaux}, \citenamefont {Perez}, \citenamefont
  {Tempo},\ and\ \citenamefont {Troncoso}}]{Bunster:2014mua}%
  \BibitemOpen
  \bibfield  {author} {\bibinfo {author} {\bibfnamefont {C.}~\bibnamefont
  {Bunster}}, \bibinfo {author} {\bibfnamefont {M.}~\bibnamefont {Henneaux}},
  \bibinfo {author} {\bibfnamefont {A.}~\bibnamefont {Perez}}, \bibinfo
  {author} {\bibfnamefont {D.}~\bibnamefont {Tempo}}, \ and\ \bibinfo {author}
  {\bibfnamefont {R.}~\bibnamefont {Troncoso}},\ }\href {\doibase
  10.1007/JHEP05(2014)031} {\bibfield  {journal} {\bibinfo  {journal} {JHEP}\
  }\textbf {\bibinfo {volume} {1405}},\ \bibinfo {pages} {031} (\bibinfo {year}
  {2014})},\ \Eprint {http://arxiv.org/abs/1404.3305} {arXiv:1404.3305
  [hep-th]} \BibitemShut {NoStop}%
\bibitem [{\citenamefont {Di~Francesco}\ \emph {et~al.}(1997)\citenamefont
  {Di~Francesco}, \citenamefont {Mathieu},\ and\ \citenamefont
  {Senechal}}]{diFrancesco}%
  \BibitemOpen
  \bibfield  {author} {\bibinfo {author} {\bibfnamefont {P.}~\bibnamefont
  {Di~Francesco}}, \bibinfo {author} {\bibfnamefont {P.}~\bibnamefont
  {Mathieu}}, \ and\ \bibinfo {author} {\bibfnamefont {D.}~\bibnamefont
  {Senechal}},\ }\href@noop {} {\emph {\bibinfo {title} {Conformal Field
  Theory}}}\ (\bibinfo  {publisher} {Springer},\ \bibinfo {year}
  {1997})\BibitemShut {NoStop}%
\bibitem [{\citenamefont {Unruh}(1976)}]{Unruh:1976db}%
  \BibitemOpen
  \bibfield  {author} {\bibinfo {author} {\bibfnamefont {W.~G.}\ \bibnamefont
  {Unruh}},\ }\href@noop {} {\bibfield  {journal} {\bibinfo  {journal} {Phys.
  Rev.}\ }\textbf {\bibinfo {volume} {D14}},\ \bibinfo {pages} {870} (\bibinfo
  {year} {1976})}\BibitemShut {NoStop}%
\bibitem [{\citenamefont {Afshar}\ \emph {et~al.}()\citenamefont {Afshar},
  \citenamefont {Detournay}, \citenamefont {Grumiller}, \citenamefont {Merbis},
  \citenamefont {Perez}, \citenamefont {Tempo},\ and\ \citenamefont
  {Troncoso}}]{prep16}%
  \BibitemOpen
  \bibfield  {author} {\bibinfo {author} {\bibfnamefont {H.}~\bibnamefont
  {Afshar}}, \bibinfo {author} {\bibfnamefont {S.}~\bibnamefont {Detournay}},
  \bibinfo {author} {\bibfnamefont {D.}~\bibnamefont {Grumiller}}, \bibinfo
  {author} {\bibfnamefont {W.}~\bibnamefont {Merbis}}, \bibinfo {author}
  {\bibfnamefont {A.}~\bibnamefont {Perez}}, \bibinfo {author} {\bibfnamefont
  {D.}~\bibnamefont {Tempo}}, \ and\ \bibinfo {author} {\bibfnamefont
  {R.}~\bibnamefont {Troncoso}},\ }\href@noop {} {}\bibinfo {note} {In
  preparation}\BibitemShut {NoStop}%
\bibitem [{\citenamefont {Cardy}(1986)}]{Cardy:1986ie}%
  \BibitemOpen
  \bibfield  {author} {\bibinfo {author} {\bibfnamefont {J.~L.}\ \bibnamefont
  {Cardy}},\ }\href@noop {} {\bibfield  {journal} {\bibinfo  {journal} {Nucl.
  Phys.}\ }\textbf {\bibinfo {volume} {B270}},\ \bibinfo {pages} {186}
  (\bibinfo {year} {1986})}\BibitemShut {NoStop}%
\bibitem [{\citenamefont {Bloete}\ \emph {et~al.}(1986)\citenamefont {Bloete},
  \citenamefont {Cardy},\ and\ \citenamefont {Nightingale}}]{Bloete:1986qm}%
  \BibitemOpen
  \bibfield  {author} {\bibinfo {author} {\bibfnamefont {H.~W.~J.}\
  \bibnamefont {Bloete}}, \bibinfo {author} {\bibfnamefont {J.~L.}\
  \bibnamefont {Cardy}}, \ and\ \bibinfo {author} {\bibfnamefont {M.~P.}\
  \bibnamefont {Nightingale}},\ }\href@noop {} {\bibfield  {journal} {\bibinfo
  {journal} {Phys. Rev. Lett.}\ }\textbf {\bibinfo {volume} {56}},\ \bibinfo
  {pages} {742} (\bibinfo {year} {1986})}\BibitemShut {NoStop}%
\bibitem [{\citenamefont {Detournay}\ \emph {et~al.}(2012)\citenamefont
  {Detournay}, \citenamefont {Hartman},\ and\ \citenamefont
  {Hofman}}]{Detournay:2012pc}%
  \BibitemOpen
  \bibfield  {author} {\bibinfo {author} {\bibfnamefont {S.}~\bibnamefont
  {Detournay}}, \bibinfo {author} {\bibfnamefont {T.}~\bibnamefont {Hartman}},
  \ and\ \bibinfo {author} {\bibfnamefont {D.~M.}\ \bibnamefont {Hofman}},\
  }\href {\doibase 10.1103/PhysRevD.86.124018} {\bibfield  {journal} {\bibinfo
  {journal} {Phys.Rev.}\ }\textbf {\bibinfo {volume} {D86}},\ \bibinfo {pages}
  {124018} (\bibinfo {year} {2012})},\ \Eprint {http://arxiv.org/abs/1210.0539}
  {arXiv:1210.0539 [hep-th]} \BibitemShut {NoStop}%
\bibitem [{\citenamefont {Cornalba}\ and\ \citenamefont
  {Costa}(2002)}]{Cornalba:2002fi}%
  \BibitemOpen
  \bibfield  {author} {\bibinfo {author} {\bibfnamefont {L.}~\bibnamefont
  {Cornalba}}\ and\ \bibinfo {author} {\bibfnamefont {M.~S.}\ \bibnamefont
  {Costa}},\ }\href {\doibase 10.1103/PhysRevD.66.066001} {\bibfield  {journal}
  {\bibinfo  {journal} {Phys.Rev.}\ }\textbf {\bibinfo {volume} {D66}},\
  \bibinfo {pages} {066001} (\bibinfo {year} {2002})},\ \Eprint
  {http://arxiv.org/abs/hep-th/0203031} {arXiv:hep-th/0203031 [hep-th]}
  \BibitemShut {NoStop}%
\bibitem [{\citenamefont {Barnich}(2012)}]{Barnich:2012xq}%
  \BibitemOpen
  \bibfield  {author} {\bibinfo {author} {\bibfnamefont {G.}~\bibnamefont
  {Barnich}},\ }\href {\doibase 10.1007/JHEP10(2012)095} {\bibfield  {journal}
  {\bibinfo  {journal} {JHEP}\ }\textbf {\bibinfo {volume} {1210}},\ \bibinfo
  {pages} {095} (\bibinfo {year} {2012})},\ \Eprint
  {http://arxiv.org/abs/1208.4371} {arXiv:1208.4371 [hep-th]} \BibitemShut
  {NoStop}%
\bibitem [{\citenamefont {Bagchi}\ \emph {et~al.}(2013)\citenamefont {Bagchi},
  \citenamefont {Detournay}, \citenamefont {Fareghbal},\ and\ \citenamefont
  {Simon}}]{Bagchi:2012xr}%
  \BibitemOpen
  \bibfield  {author} {\bibinfo {author} {\bibfnamefont {A.}~\bibnamefont
  {Bagchi}}, \bibinfo {author} {\bibfnamefont {S.}~\bibnamefont {Detournay}},
  \bibinfo {author} {\bibfnamefont {R.}~\bibnamefont {Fareghbal}}, \ and\
  \bibinfo {author} {\bibfnamefont {J.}~\bibnamefont {Simon}},\ }\href
  {\doibase 10.1103/PhysRevLett.110.141302} {\bibfield  {journal} {\bibinfo
  {journal} {Phys. Rev. Lett.}\ }\textbf {\bibinfo {volume} {110}},\ \bibinfo
  {pages} {141302} (\bibinfo {year} {2013})},\ \Eprint
  {http://arxiv.org/abs/1208.4372} {arXiv:1208.4372 [hep-th]} \BibitemShut
  {NoStop}%
\bibitem [{\citenamefont {Barnich}\ and\ \citenamefont
  {Gonzalez}(2013)}]{Barnich:2013yka}%
  \BibitemOpen
  \bibfield  {author} {\bibinfo {author} {\bibfnamefont {G.}~\bibnamefont
  {Barnich}}\ and\ \bibinfo {author} {\bibfnamefont {H.~A.}\ \bibnamefont
  {Gonzalez}},\ }\href {\doibase 10.1007/JHEP05(2013)016} {\bibfield  {journal}
  {\bibinfo  {journal} {JHEP}\ }\textbf {\bibinfo {volume} {1305}},\ \bibinfo
  {pages} {016} (\bibinfo {year} {2013})},\ \Eprint
  {http://arxiv.org/abs/1303.1075} {arXiv:1303.1075 [hep-th]} \BibitemShut
  {NoStop}%
\bibitem [{\citenamefont {Afshar}(2013)}]{Afshar:2013bla}%
  \BibitemOpen
  \bibfield  {author} {\bibinfo {author} {\bibfnamefont {H.~R.}\ \bibnamefont
  {Afshar}},\ }\href {\doibase 10.1007/JHEP10(2013)027} {\bibfield  {journal}
  {\bibinfo  {journal} {JHEP}\ }\textbf {\bibinfo {volume} {1310}},\ \bibinfo
  {pages} {027} (\bibinfo {year} {2013})},\ \Eprint
  {http://arxiv.org/abs/1307.4855} {arXiv:1307.4855 [hep-th]} \BibitemShut
  {NoStop}%
\bibitem [{\citenamefont {Gary}\ \emph {et~al.}(2015)\citenamefont {Gary},
  \citenamefont {Grumiller}, \citenamefont {Riegler},\ and\ \citenamefont
  {Rosseel}}]{Gary:2014ppa}%
  \BibitemOpen
  \bibfield  {author} {\bibinfo {author} {\bibfnamefont {M.}~\bibnamefont
  {Gary}}, \bibinfo {author} {\bibfnamefont {D.}~\bibnamefont {Grumiller}},
  \bibinfo {author} {\bibfnamefont {M.}~\bibnamefont {Riegler}}, \ and\
  \bibinfo {author} {\bibfnamefont {J.}~\bibnamefont {Rosseel}},\ }\href
  {\doibase 10.1007/JHEP01(2015)152} {\bibfield  {journal} {\bibinfo  {journal}
  {JHEP}\ }\textbf {\bibinfo {volume} {1501}},\ \bibinfo {pages} {152}
  (\bibinfo {year} {2015})},\ \Eprint {http://arxiv.org/abs/1411.3728}
  {arXiv:1411.3728 [hep-th]} \BibitemShut {NoStop}%
\bibitem [{\citenamefont {Matulich}\ \emph {et~al.}(2015)\citenamefont
  {Matulich}, \citenamefont {Perez}, \citenamefont {Tempo},\ and\ \citenamefont
  {Troncoso}}]{Matulich:2014hea}%
  \BibitemOpen
  \bibfield  {author} {\bibinfo {author} {\bibfnamefont {J.}~\bibnamefont
  {Matulich}}, \bibinfo {author} {\bibfnamefont {A.}~\bibnamefont {Perez}},
  \bibinfo {author} {\bibfnamefont {D.}~\bibnamefont {Tempo}}, \ and\ \bibinfo
  {author} {\bibfnamefont {R.}~\bibnamefont {Troncoso}},\ }\href {\doibase
  10.1007/JHEP05(2015)025} {\bibfield  {journal} {\bibinfo  {journal} {JHEP}\
  }\textbf {\bibinfo {volume} {05}},\ \bibinfo {pages} {025} (\bibinfo {year}
  {2015})},\ \Eprint {http://arxiv.org/abs/1412.1464} {arXiv:1412.1464
  [hep-th]} \BibitemShut {NoStop}%
\bibitem [{\citenamefont {Deser}\ \emph {et~al.}(1982)\citenamefont {Deser},
  \citenamefont {Jackiw},\ and\ \citenamefont {Templeton}}]{Deser:1982vy}%
  \BibitemOpen
  \bibfield  {author} {\bibinfo {author} {\bibfnamefont {S.}~\bibnamefont
  {Deser}}, \bibinfo {author} {\bibfnamefont {R.}~\bibnamefont {Jackiw}}, \
  and\ \bibinfo {author} {\bibfnamefont {S.}~\bibnamefont {Templeton}},\ }\href
  {\doibase 10.1103/PhysRevLett.48.975} {\bibfield  {journal} {\bibinfo
  {journal} {Phys. Rev. Lett.}\ }\textbf {\bibinfo {volume} {48}},\ \bibinfo
  {pages} {975} (\bibinfo {year} {1982})}\BibitemShut {NoStop}%
\bibitem [{\citenamefont {Bergshoeff}\ \emph {et~al.}(2009)\citenamefont
  {Bergshoeff}, \citenamefont {Hohm},\ and\ \citenamefont
  {Townsend}}]{Bergshoeff:2009hq}%
  \BibitemOpen
  \bibfield  {author} {\bibinfo {author} {\bibfnamefont {E.~A.}\ \bibnamefont
  {Bergshoeff}}, \bibinfo {author} {\bibfnamefont {O.}~\bibnamefont {Hohm}}, \
  and\ \bibinfo {author} {\bibfnamefont {P.~K.}\ \bibnamefont {Townsend}},\
  }\href {\doibase 10.1103/PhysRevLett.102.201301} {\bibfield  {journal}
  {\bibinfo  {journal} {Phys. Rev. Lett.}\ }\textbf {\bibinfo {volume} {102}},\
  \bibinfo {pages} {201301} (\bibinfo {year} {2009})},\ \Eprint
  {http://arxiv.org/abs/0901.1766} {arXiv:0901.1766 [hep-th]} \BibitemShut
  {NoStop}%
\bibitem [{\citenamefont {Deser}(2009)}]{Deser:2009hb}%
  \BibitemOpen
  \bibfield  {author} {\bibinfo {author} {\bibfnamefont {S.}~\bibnamefont
  {Deser}},\ }\href {\doibase 10.1103/PhysRevLett.103.101302} {\bibfield
  {journal} {\bibinfo  {journal} {Phys. Rev. Lett.}\ }\textbf {\bibinfo
  {volume} {103}},\ \bibinfo {pages} {101302} (\bibinfo {year} {2009})},\
  \Eprint {http://arxiv.org/abs/0904.4473} {arXiv:0904.4473 [hep-th]}
  \BibitemShut {NoStop}%
\bibitem [{\citenamefont {Bergshoeff}\ \emph {et~al.}(2013)\citenamefont
  {Bergshoeff}, \citenamefont {de~Haan}, \citenamefont {Hohm}, \citenamefont
  {Merbis},\ and\ \citenamefont {Townsend}}]{Bergshoeff:2013xma}%
  \BibitemOpen
  \bibfield  {author} {\bibinfo {author} {\bibfnamefont {E.~A.}\ \bibnamefont
  {Bergshoeff}}, \bibinfo {author} {\bibfnamefont {S.}~\bibnamefont {de~Haan}},
  \bibinfo {author} {\bibfnamefont {O.}~\bibnamefont {Hohm}}, \bibinfo {author}
  {\bibfnamefont {W.}~\bibnamefont {Merbis}}, \ and\ \bibinfo {author}
  {\bibfnamefont {P.~K.}\ \bibnamefont {Townsend}},\ }\href {\doibase
  10.1103/PhysRevLett.111.111102, 10.1103/PhysRevLett.111.259902} {\bibfield
  {journal} {\bibinfo  {journal} {Phys.Rev.Lett.}\ }\textbf {\bibinfo {volume}
  {111}},\ \bibinfo {pages} {111102} (\bibinfo {year} {2013})},\ \Eprint
  {http://arxiv.org/abs/1307.2774} {arXiv:1307.2774} \BibitemShut {NoStop}%
\bibitem [{\citenamefont {Afshar}\ \emph {et~al.}(2011)\citenamefont {Afshar},
  \citenamefont {Cvetkovic}, \citenamefont {Ertl}, \citenamefont {Grumiller},\
  and\ \citenamefont {Johansson}}]{Afshar:2011yh}%
  \BibitemOpen
  \bibfield  {author} {\bibinfo {author} {\bibfnamefont {H.}~\bibnamefont
  {Afshar}}, \bibinfo {author} {\bibfnamefont {B.}~\bibnamefont {Cvetkovic}},
  \bibinfo {author} {\bibfnamefont {S.}~\bibnamefont {Ertl}}, \bibinfo {author}
  {\bibfnamefont {D.}~\bibnamefont {Grumiller}}, \ and\ \bibinfo {author}
  {\bibfnamefont {N.}~\bibnamefont {Johansson}},\ }\href {\doibase
  10.1103/PhysRevD.84.041502} {\bibfield  {journal} {\bibinfo  {journal}
  {Phys.Rev.}\ }\textbf {\bibinfo {volume} {D84}},\ \bibinfo {pages}
  {041502(R)} (\bibinfo {year} {2011})},\ \Eprint
  {http://arxiv.org/abs/1106.6299} {arXiv:1106.6299 [hep-th]} \BibitemShut
  {NoStop}%
\bibitem [{\citenamefont {Afshar}\ \emph {et~al.}(2012)\citenamefont {Afshar},
  \citenamefont {Cvetkovic}, \citenamefont {Ertl}, \citenamefont {Grumiller},\
  and\ \citenamefont {Johansson}}]{Afshar:2011qw}%
  \BibitemOpen
  \bibfield  {author} {\bibinfo {author} {\bibfnamefont {H.}~\bibnamefont
  {Afshar}}, \bibinfo {author} {\bibfnamefont {B.}~\bibnamefont {Cvetkovic}},
  \bibinfo {author} {\bibfnamefont {S.}~\bibnamefont {Ertl}}, \bibinfo {author}
  {\bibfnamefont {D.}~\bibnamefont {Grumiller}}, \ and\ \bibinfo {author}
  {\bibfnamefont {N.}~\bibnamefont {Johansson}},\ }\href@noop {} {\bibfield
  {journal} {\bibinfo  {journal} {Phys.Rev.}\ }\textbf {\bibinfo {volume}
  {D85}},\ \bibinfo {pages} {064033} (\bibinfo {year} {2012})},\ \Eprint
  {http://arxiv.org/abs/1110.5644} {arXiv:1110.5644 [hep-th]} \BibitemShut
  {NoStop}%
\bibitem [{\citenamefont {Henneaux}\ and\ \citenamefont
  {Rey}(2010)}]{Henneaux:2010xg}%
  \BibitemOpen
  \bibfield  {author} {\bibinfo {author} {\bibfnamefont {M.}~\bibnamefont
  {Henneaux}}\ and\ \bibinfo {author} {\bibfnamefont {S.-J.}\ \bibnamefont
  {Rey}},\ }\href {\doibase 10.1007/JHEP12(2010)007} {\bibfield  {journal}
  {\bibinfo  {journal} {JHEP}\ }\textbf {\bibinfo {volume} {1012}},\ \bibinfo
  {pages} {007} (\bibinfo {year} {2010})},\ \Eprint
  {http://arxiv.org/abs/1008.4579} {arXiv:1008.4579 [hep-th]} \BibitemShut
  {NoStop}%
\bibitem [{\citenamefont {Campoleoni}\ \emph {et~al.}(2010)\citenamefont
  {Campoleoni}, \citenamefont {Fredenhagen}, \citenamefont {Pfenninger},\ and\
  \citenamefont {Theisen}}]{Campoleoni:2010zq}%
  \BibitemOpen
  \bibfield  {author} {\bibinfo {author} {\bibfnamefont {A.}~\bibnamefont
  {Campoleoni}}, \bibinfo {author} {\bibfnamefont {S.}~\bibnamefont
  {Fredenhagen}}, \bibinfo {author} {\bibfnamefont {S.}~\bibnamefont
  {Pfenninger}}, \ and\ \bibinfo {author} {\bibfnamefont {S.}~\bibnamefont
  {Theisen}},\ }\href {\doibase 10.1007/JHEP11(2010)007} {\bibfield  {journal}
  {\bibinfo  {journal} {JHEP}\ }\textbf {\bibinfo {volume} {1011}},\ \bibinfo
  {pages} {007} (\bibinfo {year} {2010})},\ \Eprint
  {http://arxiv.org/abs/1008.4744} {arXiv:1008.4744 [hep-th]} \BibitemShut
  {NoStop}%
\bibitem [{\citenamefont {Gaberdiel}\ and\ \citenamefont
  {Hartman}(2011)}]{Gaberdiel:2011wb}%
  \BibitemOpen
  \bibfield  {author} {\bibinfo {author} {\bibfnamefont {M.~R.}\ \bibnamefont
  {Gaberdiel}}\ and\ \bibinfo {author} {\bibfnamefont {T.}~\bibnamefont
  {Hartman}},\ }\href {\doibase 10.1007/JHEP05(2011)031} {\bibfield  {journal}
  {\bibinfo  {journal} {JHEP}\ }\textbf {\bibinfo {volume} {1105}},\ \bibinfo
  {pages} {031} (\bibinfo {year} {2011})},\ \Eprint
  {http://arxiv.org/abs/1101.2910} {arXiv:1101.2910 [hep-th]} \BibitemShut
  {NoStop}%
\bibitem [{\citenamefont {Afshar}\ \emph {et~al.}(2013)\citenamefont {Afshar},
  \citenamefont {Bagchi}, \citenamefont {Fareghbal}, \citenamefont
  {Grumiller},\ and\ \citenamefont {Rosseel}}]{Afshar:2013vka}%
  \BibitemOpen
  \bibfield  {author} {\bibinfo {author} {\bibfnamefont {H.}~\bibnamefont
  {Afshar}}, \bibinfo {author} {\bibfnamefont {A.}~\bibnamefont {Bagchi}},
  \bibinfo {author} {\bibfnamefont {R.}~\bibnamefont {Fareghbal}}, \bibinfo
  {author} {\bibfnamefont {D.}~\bibnamefont {Grumiller}}, \ and\ \bibinfo
  {author} {\bibfnamefont {J.}~\bibnamefont {Rosseel}},\ }\href {\doibase
  10.1103/PhysRevLett.111.121603} {\bibfield  {journal} {\bibinfo  {journal}
  {Phys.Rev.Lett.}\ }\textbf {\bibinfo {volume} {111}},\ \bibinfo {pages}
  {121603} (\bibinfo {year} {2013})},\ \Eprint {http://arxiv.org/abs/1307.4768}
  {arXiv:1307.4768 [hep-th]} \BibitemShut {NoStop}%
\bibitem [{\citenamefont {Gonzalez}\ \emph {et~al.}(2013)\citenamefont
  {Gonzalez}, \citenamefont {Matulich}, \citenamefont {Pino},\ and\
  \citenamefont {Troncoso}}]{Gonzalez:2013oaa}%
  \BibitemOpen
  \bibfield  {author} {\bibinfo {author} {\bibfnamefont {H.~A.}\ \bibnamefont
  {Gonzalez}}, \bibinfo {author} {\bibfnamefont {J.}~\bibnamefont {Matulich}},
  \bibinfo {author} {\bibfnamefont {M.}~\bibnamefont {Pino}}, \ and\ \bibinfo
  {author} {\bibfnamefont {R.}~\bibnamefont {Troncoso}},\ }\href {\doibase
  10.1007/JHEP09(2013)016} {\bibfield  {journal} {\bibinfo  {journal} {JHEP}\
  }\textbf {\bibinfo {volume} {1309}},\ \bibinfo {pages} {016} (\bibinfo {year}
  {2013})},\ \Eprint {http://arxiv.org/abs/1307.5651} {arXiv:1307.5651
  [hep-th]} \BibitemShut {NoStop}%
\bibitem [{\citenamefont {Barnich}\ and\ \citenamefont
  {Troessaert}(2010)}]{Barnich:2009se}%
  \BibitemOpen
  \bibfield  {author} {\bibinfo {author} {\bibfnamefont {G.}~\bibnamefont
  {Barnich}}\ and\ \bibinfo {author} {\bibfnamefont {C.}~\bibnamefont
  {Troessaert}},\ }\href {\doibase 10.1103/PhysRevLett.105.111103} {\bibfield
  {journal} {\bibinfo  {journal} {Phys. Rev. Lett.}\ }\textbf {\bibinfo
  {volume} {105}},\ \bibinfo {pages} {111103} (\bibinfo {year} {2010})},\
  \Eprint {http://arxiv.org/abs/0909.2617} {arXiv:0909.2617 [gr-qc]}
  \BibitemShut {NoStop}%
\end{thebibliography}

%

\end{document}